\documentclass[a4paper,11pt]{article}
\pdfoutput=1 

\usepackage{jheppub_BLNK} 

\usepackage[T1]{fontenc} 
\usepackage{float}
\usepackage{epstopdf}
\usepackage{amsmath}

\title{\boldmath Lovelock black holes surrounded by dark fluid in power-Yang-Mills massive gravity}

\author{Askar Ali and Khalid Saifullah} 
\affiliation{Department of Mathematics, Quaid-i-Azam University, Islamabad, Pakistan}

\emailAdd{askarali@math.qau.edu.pk} \emailAdd{ksaifullah@fas.harvard.edu}

\abstract{We consider a model where massive static spherically symmetric black hole, in the presence of power-Yang-Mills source, is surrounded by a dark fluid with non-linear equation of state. In this set up we construct a new class of magnetized Lovelock black hole solutions of the gravitational field equations. In particular, we work out the metric functions in both D-dimensional massive Einstein and massive Gauss-Bonnet gravities. We study thermodynamics of these black holes also and show that the mass and associated thermodynamic quantities like Hawking temperature and heat capacity depend on parameters of the dark fluid and the power-Yang-Mills magnetic source. We note that the entropy does not satisfy the area law in Gauss-Bonnet and higher order Lovelock black holes. Furthermore, phase transitions of black holes in each case are also discussed.
\vspace{80 mm}
}

\notoc 

\begin{document}
\maketitle
\flushbottom 


\section{Motivation}
\label{sec:intro}
Einstein's theory of gravity has already been tested at low energy scales \cite{1}. However, at high energy close to the Planck scale, it is largely expected that this theory must be modified. String theory and brane cosmology strongly predict that there also exists the phenomenon of higher dimensions and so it is worthwhile to think about the generalization of gravity into higher dimensions \cite{2,2a,3,3a,3b}. In this context Lovelock \cite{4} introduced a gravity theory which involves the addition of high curvature terms to the familiar Einstein-Hilbert action in higher dimensions. It is worth-mentioning that this theory does not include derivative terms higher than the second derivatives of the metric and so it is a ghost-free theory \cite{5}. 

Four-dimensional black hole solutions of both the Einstein and Einstein-Maxwell equations have a long history with higher dimensional versions of such black holes also studied with interest during the last several decades \cite{5a}. More recently charged black hole solution has also been found in which the source of gravity is a non-linear electromagnetic field and the non-linearity is defined in the powers of Maxwell's invariant $(F_{\mu\nu}F^{\mu\nu})^q$, where $q$ is a real number \cite{6}. The non-linearity involved in this power-Maxwell theory is fundamentally different from the well-known Born-Infeld electrodynamics \cite{7}. This different behaviour is seen from definition of the Lagrangian density associated with the Born-Infeld electromagnetic field which yields the nonzero trace of the energy-momentum tensor. However, in the power-Maxwell field when one considers the special case $q=D/4$, where $D$ is the dimension of spacetime, it bears the traceless energy-momentum tensor meaning that the principle of conformal invariance is satisfied in this formalism. Since black holes of gravity coupled to Maxwell electrodynamics are familiar, one is inspired to consider the extension of such coupling into the study of black holes in the presence of Yang-Mills field which is relatively new \cite{7a,7b,7c}. The Yang-Mills field, unlike Maxwell's electrodynamics, generates a complicated structure which makes it difficult to find solutions to this system. In spite of complications, when we consider spherical symmetry and special gauge group it is possible to obtain the analytical solutions. In this regard black hole solutions have been found where Yang-Mills field is taken as a source of gravitational field \cite{7b,7c,7d,7e}. These black hole solutions are obtained with the help of an alternate technique i.e. by using Wu-Yang ansatz in higher dimensions \cite{7f,7g}. Instead of assuming the Yang-Mills theory, it is also possible to explore black hole spacetimes with power-Yang-Mills sources \cite{8}, i.e., one can pick the source of gravity as $(F^{(a)}_{\mu\nu}F^{(a)\mu\nu})^q$, with $F^{(a)}_{\mu\nu}$ being the Yang-Mills field where $1\leq a\leq (D-1)(D-2)/2$ and $q$ is a real number. Here, these Lovelock black holes have been investigated and magnetically charged solutions are constructed. Similarly, different types of magnetized Lovelock black hole solutions of dimensionally continued gravity have also been found recently \cite{9}.

On the observational front a lot of data clearly shows that there exists accelerated expansion of the Universe. It is widely believed that perhaps this accelerated expansion is due to the presence of gravitationally self-repulsive dark energy. From the 2018 release of the Planck data related to CMB power spectra and CMB lensing, when seen in conjunction with BAO observations, it is concluded that baryonic matter component is not greater than 5 percent of the total energy density, and that the contribution of invisible dark energy and dark matter in the Universe is about 95 percent of the energy density \cite{10}. The dominant contribution of the invisible dark sector to the Universe shows the importance of studying massive gravitating objects in the presence of these types of mysterious exotic fields. Quintessence, as a possible candidate for this invisible dark sector, is specified by the linear equation of state $p_{\overline{q}}=w\rho_{\overline{q}}$, where $p_{\overline{q}}$ represents the pressure, $\rho_{\overline{q}}$ is the energy density and the parameter $w$ is defined by the inequality $-1<w<-1/3$. Many spherically symmetric black hole solutions have been extensively discussed in the literature \cite{11,12,13,14,15,16,18,19,20,20a}, among which Refs. \cite{18,19,20,20a} mainly focus on Lovelock black holes. Speaking of the universal dark sector, it is also possible that the energy component is a hybrid of dark matter and dark energy known as the unified dark fluid. An example of such a unified dark fluid is the Chaplygin gas \cite{21} whose generalized model \cite{22,23} has also been studied in an attempt to understand the accelerated expansion of the Universe \cite{24,25,26}. This hybrid dark fluid has a very amazing property that its dynamical behaviour is like dark matter in the initial times but it appears like dark energy in late times. So as to attain this behaviour, the dark fluid is considered to satisfy an exotic non-linear equation of state. The static spherically symmetric black hole solutions of Lovelock gravity have been derived \cite{27} with electrostatic field and having a Chaplygin-like dark fluid described by the equation of state $p_d=-B/\rho_d$ as a source of gravitational field. Here index $d$ represents the dark fluid, and $B$ is a positive constant. 

In this paper we assume the matter contents in the form of power-Yang-Mills field and dark fluid, and consider the contributions of massive graviton in our formalism so that the Lovelock gravity is modified into a massive Lovelock theory. Although graviton is massless in the Einstein theory but the question arises if it is possible to construct a self-consistent gravity theory with massive graviton, and to claim to modify the Einstein gravity and come up with a massive gravity. From the theoretical perspective, the shear difficulties and complications in the construction of massive gravity make this subject more interesting. In this context the ghost-free theory of free massive gravitons was developed by Fierz and Pauli \cite{27a,27b} where the massive gravitons were assumed to be non-interactive in flat background. It was also suggested that extension of this theory to curved spacetimes would lead to the possibilities of ghost instabilities \cite{27c}. The massive gravity and its nonlinear extension in the absence of ghost field in quantum regime \cite{27d,27e}, where the effects of quantum interactions are taken into account, have also been investigated \cite{27f,27g,27h}. The influence of massive graviton on gravity waves during the inflation era has also been discussed \cite{27i}. In addition, the contributions and effects of massive gravity have also been discussed for astrophysical objects, for example, the maximum mass of neutron stars greater than $3M_{sun}$ \cite{28} has been calculated in massive gravity. Furthermore, a special class of charged massive black holes \cite{28a,28b,28c,28d,28e,28f,28g} and a family of non-trivial massive black holes in AdS background have also been investigated \cite{28h}. Massive gravity also enables modification in thermal stability of black holes \cite{29,30,31,32,33,34,35}. In particular, the van der Waals-like behaviour in the case of non-spherical black holes \cite{36}, remnant of Hawking temperature \cite{37} and the phenomenon of anti-evaporation have also been studied in the literature \cite{38,39}. It may be noted that the behaviour of massive graviton in massive gravity is similar to that of lattice in holographic conductor model. Recently, thermodynamic stability and P-V criticality of  Vegh’s massive gravity has been studied \cite{39a,39b}. Some holographic consequences related to the mass of graviton have also been investigated \cite{39c,39d,39e}. In addition to black hole solutions of massive gravity, charged black hole solutions of Gauss-Bonnet massive gravity and their thermodynamics have also been found recently \cite{40}.

Now, the simplest way to come up with a massive gravity theory is by adding the mass term in the Einstein-Hilbert action, and recover Einstein's theory in the limit $m\rightarrow 0$, where $m$ is the mass of graviton. Thus, in this paper, we intend to generalize the action of Lovelock gravity by adding the mass term and constructing black hole solutions of massive Lovelock gravity. Generally, the higher derivative gravity terms (such as Lovelock's) are corrections that are significant at high energy regime whereas the mass terms (of massive gravity) play their role at low energy. We are not considering the energy regime, directly, and analyse the weight of Lovelock (Gauss-Bonnet and Einstein, in particular) and mass terms contributions for the possibility of new black hole solutions in this theory. In particular, we consider the power-Yang-Mills field as a source of gravitational field and construct a new class of Lovelock black hole solutions of massive gravity in the presence of Chaplygin-like dark fluid having a non-linear equation of state. We also study thermodynamics of these types of Lovelock black holes and work out different thermodynamic quantities associated with black holes.

The paper is constructed as follows. In the next section, we provide a brief introduction to Lovelock gravity and perform its coupling with the power-Yang-Mills field theory. We also include the effects of massive gravity in this context and construct a family of magnetized Lovelock black hole solutions where we assume that these black holes are surrounded by a dark fluid. In this section the gravitational field equations are solved and different thermodynamic quantities associated to Lovelock black holes are calculated. In Sections 3 and 4, black holes of massive gravity and Gauss-Bonnet gravity have been studied, respectively. Finally, we summarize and conclude the paper in Section 5.

\section{Magnetized Lovelock black holes of massive gravity} 

The action function for Lovelock gravity with matter sources in diverse dimensions can be written in the form
\begin{equation}
I=I_L+I_M,
\label{1}
\end{equation}
where $I_M$ corresponds to the action function of matter contents in the spacetime i.e. the power-Yang-Mills field and dark fluid. In this paper we want to obtain the Lovelock black hole solution of massive power-Yang-Mills gravity surrounded by dark fluid. The Lovelock action $I_L$ is given by 
\begin{equation}
I_L=\frac{1}{2}\int d^Dx\sqrt{-g}\bigg[\sum_{p=0}^{s}\frac{\alpha_p}{2^p}\delta^{\mu_1...\mu_{2p}}_{\nu_1...\nu_{2p}} R^{\nu_1\nu_2}_{\mu_1\mu_2}...R^{\nu_{2p-1}\nu_{2p}}_{\mu_{2p-1}\mu_{2p}}+m^2\sum_{i=1}^{4}c_iU_i(\textbf{g},\textbf{f})\bigg],
\label{2}
\end{equation}
where $\textbf{g}$ is the metric tensor, $\textbf{f}$ is the fixed symmetric tensor, $m$ is the massive parameter which represents the mass of graviton, $R^{\alpha\beta}_{\mu\nu}$ represent the curvature tensor components and $\delta^{\mu_1...\mu_{2p}}_{\nu_1...\nu_{2p}}$ is the generalized Kronecker delta of order $2p$ and $s=(D-1)/2$ where $s$ is a natural number. The coefficients $\alpha_p$ in (\ref{1}) are arbitrary constants in which $\alpha_0=-2\Lambda$ means that it is related to the cosmological constant. Furthermore, $c_i$'s are constants and $U_i$'s are the symmetric polynomials of eigenvalues of the $D\times D$ matrix $K^{\mu}_{\nu}=\sqrt{g^{\mu\alpha}f_{\alpha\nu}}$ which can be obtained as
\begin{eqnarray}\begin{split}
U_1=[K], U_2=[K]^2-[K^2], 
U_3=[K]^3-3[K][K^2]+2[K^3],\\
U_4=[K]^4-6[K^2][K]^2+8[K^3][K]+3[K^2]^2-6[K^4]. \label{3}\end{split}
\end{eqnarray}
 Note that, the reference metric can be defined as $f_{\mu\nu}=diag(0,0,\epsilon^2r^2h_{ij})$, where $\epsilon$ is a positive constant. So, by using this reference metric we can determine \cite{39a} the polynomials $U_i$ in the form $U_j=\frac{\epsilon^j}{r^j}\prod_{x=2}^{j+1}(d-x)$.

If we take the variation of action (\ref{1}) with respect to the metric tensor, $g_{\mu\nu}$, we obtain the equations of the gravitational field as 
\begin{equation}
\sum_{p=0}^{s}\frac{\alpha_p}{2^{p+1}}\delta^{\nu\lambda_1...\lambda_{2p}}_{\mu\rho_1...\rho_{2p}} R^{\rho_1\rho_2}_{\lambda_1\lambda_2}...R^{\rho_{2p-1}\rho_{2p}}_{\lambda_{2p-1}\lambda_{2p}}+m^2X^{\nu}_{\mu}=T^{\nu}_{\mu},
\label{4}
\end{equation}
where $T^{\nu}_{\mu}$ represents the stress-energy tensor given by
\begin{equation}
T_{\mu\nu}=-\frac{2}{\sqrt{-g}}\frac{\delta I_M}{\delta g^{\mu\nu}},
\label{5}
\end{equation}
and the tensor components $X^{\nu}_{\mu}$ related to the contributions from massive gravity are given by
\begin{eqnarray}
\label{5a}
\begin{split}
X_{\mu\nu}&=-\frac{c_1}{2}\bigg(U_1g_{\mu\nu}-K_{\mu\nu}\bigg)-\frac{c_2}{2}\bigg(U_2g_{\mu\nu}-2U_1K_{\mu\nu}+2K^2_{\mu\nu}\bigg)\\&-\frac{c_3}{2} \bigg(U_3g_{\mu\nu}-3U_2K_{\mu\nu}+6U_1K^2_{\mu\nu}-6K^3_{\mu\nu}\bigg)\\&-\frac{c_4}{2}\bigg(U_4g_{\mu\nu}-4U_3K_{\mu\nu}+2U_2K^2_{\mu\nu}-24U_1K^3_{\mu\nu}+24K^4_{\mu\nu}\bigg).
\end{split}
\end{eqnarray}
Now, the general static and spherically symmetric line element in $D$-dimensions can be expressed as 
\begin{equation}
ds^2=-f(r)dt^2+\frac{dr^2}{f(r)}+r^2 (h_{ij}dx^idx^j),
\label{6}
\end{equation}
where $h_{ij}dx^idx^j$ is the metric of $D-2$ dimensional hypersurface having constant scalar curvature $k$ such that the values $k=1,0,-1$, represent spherical, flat and hyperbolic spaces, respectively.

Now the Lagrangian density corresponding to power-Yang-Mills field is given by
 \begin{equation}
L_{PYM}=-(\Upsilon)^q,        \label{7}
\end{equation}
where $\Upsilon$ denotes the Yang-Mills invariant given by
\begin{equation}
\Upsilon=\sum_{a=1}^{(D-2)(D-1)/2}\bigg(F^{a}_{\lambda\sigma}F^{(a)\lambda\sigma}\bigg), \label{8}
\end{equation}
$q$ is a positive real parameter and the Yang-Mills field is defined by
\begin{equation}
F^{(a)}=dA^{(a)}+\frac{1}{2\eta}C^{(a)}_{(b)(c)}A^{(b)}\wedge A^{(c)},\label{9}
\end{equation}
where, $C^{(a)}_{(b)(c)}$ are the structure constants of $(D-2)(D-1)/2$ -parameter Lie group G, $\eta$ represents the coupling constant and $A^{(a)}$ are Yang-Mills potentials of the $So(D-1)$ gauge group. These structure constants have been calculated in Ref. \cite{46}. It should be noted that it does not matter whether we write the internal indices $[a,b,...]$ in covariant or contravariant form. Thus, using the Lagrangian density (\ref{7}), the stress-energy tensor of power-Yang-Mills field is given by
\begin{equation}
T^{(a)\nu}_{\mu}=-\frac{1}{2}\bigg[\delta^{\nu}_{\mu}\Upsilon^q-4q \sum_{a=1}^{(D-2)(D-1)/2}\bigg(F^{(a)}_{\mu\lambda}F^{(a)\mu\lambda}\bigg)\Upsilon^{q-1}\bigg].\label{10}
\end{equation}
We obtain the equations of Yang-Mills field by varying the action (\ref{1}) with respect to the gauge potentials $A^{(a)}$
\begin{equation} 
d(^{\star}F^{(a)}\Upsilon^{q-1})+\frac{1}{\eta}C^{(a)}_{(b)(c)}\Upsilon^{q-1}A^{(b)}\wedge^{\star}F^{(c)}=0,  \label{11}
\end{equation}
where $\star$ is used to denote the duality of a field. If we take the static metric given in (\ref{6}) and using the Wu-Yang ansatz in Einstein-Yang-Mills theory of gravity \cite{7f,7g}, then the power-Yang-Mills equations \cite{47,48} are satisfied if the Yang-Mills magnetic gauge potential one-forms in this ansatz are described by 
\begin{equation}
A^{(a)}=\frac{Q}{r^2}C^{(a)}_{(i)(j)}x^{i}dx^{j}, r^2=\sum_{i=1}^{d-1}x_{i}^2,   \label{12}
\end{equation}
where the constant $Q$ is associated with the Yang-Mills magnetic charge and $2\leq j+1\leq i\leq d-1$. Therefore, the symmetric stress-energy tensor (\ref{41}), with
\begin{equation}
\Upsilon=\frac{n(n-1)Q^2}{r^4}=\frac{(d-2)(d-3)Q^2}{r^4}, \label{13}
\end{equation}
becomes
\begin{equation}
T^{(a)\nu}_{\mu}=-\frac{1}{2}\Upsilon^q diag[1,1,\zeta,\zeta,...,\zeta], \zeta=\bigg(1-\frac{4q}{(D-2)}\bigg).\label{14}
\end{equation}
Now we consider the dark fluid with non-linear equation of state $p_d=-B/\rho_d$, where $B$ is a constant. For spherically symmetric $D$-dimensional spacetime geometry, the stress-energy tensor corresponding to the dark fluid can be defined as
\begin{eqnarray}\begin{split}
T^{(d) t}_{t}=\Pi(r),   T^{(d) i}_{i}=0, 
  T^{(d) j}_{i}=\Delta(r)\frac{r_ir^j}{r_nr^n}+\Theta(r)\delta^{j}_{i}.   \label{15}\end{split}
\end{eqnarray}
If we consider isotropic average over the angles
\begin{equation}
[r_ir^j]_{A}=\frac{r_nr^n \delta^{j}_{i}}{D-1},\label{16}
\end{equation}
we can get
\begin{equation}
[T^{(d) j}_{i}]_{A}=\bigg(\frac{\Delta(r)}{D-1}+\Theta(r)\bigg)\delta^{j}_{i}=p_d(r)\delta^{j}_{i}=-\frac{B}{\rho_d(r)}.\label{17}
\end{equation}
Using the condition of staticity and spherical symmetry, $T^{(d) t}_{t}=T^{(d) r}_{r}=-\rho_d(r)$, in (\ref{17}) we have the following expressions 
  \begin{eqnarray}\begin{split}
  \Pi(r)&=-\rho_d(r), \\
  \Delta (r)&=-\frac{(D-1)\rho_d}{(D-2)}+\frac{(D-1)B}{(D-2)\rho_d},\\ \Theta(r)&=\frac{\rho_d}{(D-2)}-\frac{(D-1)B}{(D-2)\rho_d}. \label{18}\end{split}
  \end{eqnarray}
  So, the other angular components of the stress-energy tensor are calculated as
   \begin{eqnarray}\begin{split}
   T^{(d) \theta_1}_{\theta_1}=T^{(d) \theta_2}_{\theta_2}=\cdots=T^{(d) \theta_{D-2}}_{\theta_{D-2}}=\frac{\rho_d(r)}{D-2}-\frac{(D-1)B}{(D-2)\rho_d(r)}.\label{19}\end{split}
   \end{eqnarray}
   Let us introduce the metric function $f(r)$ in line element (\ref{6}) as
   \begin{equation}
   f(r)=k-r^2V(r),\label{20}
   \end{equation}
   where the function $V(r)$ is related to the polynomial
   \begin{equation}
   P[V(r)]=\sum_{p=0}^{s}\overline{\alpha}_pV^p(r),\label{21}
   \end{equation}
   whose coefficients $\overline{\alpha}_p$ are defined as
   \begin{eqnarray}\begin{split}
   \overline{\alpha}_0&=\frac{\alpha_0}{(D-1)(D-2)},              \overline{\alpha}_1=1, \\
   \overline{\alpha}_p&=\prod_{i=3}^{2p}(D-i)\alpha_p.  \label{22}\end{split}
   \end{eqnarray}
   The general expression for $\overline{\alpha}_p$ in (\ref{22}) holds only for $p>3$. By using the above components of the stress-energy tensor,  $T_{\mu\nu}=T^{(a)}_{\mu\nu}+T^{(d)}_{\mu\nu}$, in the Lovelock field equations (\ref{4}) we obtain the differential equation for energy density of the dark fluid in the form
   \begin{equation}
   \frac{d\rho_d}{dr}+\frac{(D-1)}{r}\rho_d=\frac{B(D-1)}{r\rho_d},\label{23}
   \end{equation}
    which can be solved to give
    \begin{equation}
    \rho_d(r)=\sqrt{B+\frac{S^2}{r^{2D-2}}},\label{24}
    \end{equation}
where, $S>0$ is the constant of integration. It is clearly seen that this is valid in any particular Lovelock gravity theory and the energy density of the dark fluid will still satisfy the same differential equation (\ref{23}). Furthermore, we can see from Eq. (\ref{24}) that in the limit $r\rightarrow\infty$, $B$ approaches the value $\sqrt{B}$, which implies that at large distances from the gravitating object the dark fluid behaves like a cosmological constant. Similarly, when we use the above information of matter sources and the line element (\ref{6}) in gravitational field equations (\ref{4}) we can obtain the polynomial equation for the magnetized black hole as
    \begin{eqnarray}\begin{split}
    P[V(r)]&=\frac{2M}{(D-2)\Sigma_{D-2}r^{D-1}}+\frac{(D-2)^{q-1}(D-3)^qQ^{2q}}{(D-4q-1)r^{4q}} \\ &+\frac{2r^{1-D}}{(D-2)(D-1)}\bigg[r^{D-1}\sqrt{B+\frac{S^2}{r^{2D-2}}}-S\arcsin{\bigg(\frac{S}{\sqrt{B}r^{D-1}}\bigg)}\bigg] \\&-m^2\bigg[\frac{c_1\epsilon}{r}+\frac{c_2\epsilon^2(D-2)}{(D-3)r^2}+\frac{c_3\epsilon^3(D-3)}{r^3}+\frac{c_4\epsilon^4(D-3)(D-4)}{r^4}\bigg], \label{25}\end{split}
    \end{eqnarray}
    which is satisfied by function $V(r)$ and hence by the metric function $f(r)$ as well. Note that $M$ is a constant of integration associated with the black hole mass and
    \begin{equation}
    \Sigma_{D-2}=\frac{2\pi^{\frac{(D-1)}{2}}}{\Gamma{\bigg(\frac{D-1}{2}\bigg)}},\label{25a}
    \end{equation}
      is the volume of $D-2$-dimensional hypersurface. Next we will discuss thermodynamics of Lovelock black holes generated by polynomial equation (\ref{25}). For this we will calculate different thermodynamic quantities for the black hole in terms of the event horizon $r_+$ which satisfies the equation $f(r_+)=0$. So using Eq. (\ref{20}) we can write
    \begin{equation}
    r_+^2=\frac{k}{V(r_+)}.\label{26}
    \end{equation}
    Thus, the total mass in terms of horizon's radius $r_+$ is given by 
    \begin{eqnarray}\begin{split}
    M&=\frac{\Sigma_{D-2}}{2}\bigg[\sum_{p=0}^{s}\frac{\overline{\alpha}_pk^p(D-2)}{r_+^{-(D-2p-1)}}-\frac{(D-2)^{q}(D-3)^qQ^{2q}}{(D-4q-1)r_+^{4q-D+1}}+m^2\bigg(\frac{c_1\epsilon(D-2)}{r_+^{2-D}}\\&+\frac{c_2\epsilon^2(D-2)^2}{(D-3)r_+^{3-D}}+\frac{c_3\epsilon^3(D-2)(D-3)}{r_+^{4-D}}+\frac{c_4\epsilon^4(D-2)(D-3)(D-4)}{r_+^{5-D}}\bigg)\\&-\frac{2}{D-1}\bigg(r_+^{D-1}\sqrt{B+\frac{S^2}{r_+^{2D-2}}}-S\arcsin{\bigg(\frac{S}{\sqrt{B}r_+^{D-1}}\bigg)}\bigg)\bigg]. \label{27}\end{split}
    \end{eqnarray}
    The Hawking temperature of Lovelock black holes governed by the polynomial equation (\ref{25}) is defined in terms of surface gravity $\kappa$ as $T=\kappa/2\pi$. Thus, using the definition of surface gravity yields the Hawking temperature
     \begin{eqnarray}\begin{split}
   T&=\frac{1}{4\pi W(r_+)}\bigg[\sum_{p=0}^{s}\frac{\overline{\alpha}_pk^p(D-2p-1)}{r_+^{2p+1}}-\frac{(D-2)^{q-1}(D-3)^qQ^{2q}}{r_+^{4q+1}} \\&+m^2\bigg(\frac{c_1\epsilon(D-2)}{r_+^{2}}+\frac{c_2\epsilon^2(D-2)}{r_+^{3}}+\frac{c_3\epsilon^3(D-3)(D-4)}{r_+^{4}} \\&+\frac{c_4\epsilon^4(D-5)(D-3)(D-4)}{r_+^{5}}\bigg)-\frac{2}{(D-2)r_+}\sqrt{B+\frac{S^2}{r_+^{2D-2}}}\bigg], \label{28}\end{split} 
    \end{eqnarray}
    where $W(r_+)$ is defined by
    \begin{equation}
    W(r_+)=\sum_{p=0}^{s}\frac{p\overline{\alpha}_pk^{p-1}}{r_+^{2p}}.\label{29}
    \end{equation}
    The entropy of the black hole is given by
    \begin{equation}
    S=\int \frac{dM}{T}=\int T^{-1}\frac{dM}{dr_+}dr_+. \label{30}
    \end{equation}
    Now, differentiating the mass (\ref{27}) with respect to $r_+$ we obtain
    \begin{eqnarray}\begin{split}
    \frac{dM}{dr_+}&=\frac{\Sigma_{D-2}}{2}\bigg[\sum_{p=0}^{s}\frac{\overline{\alpha}_pk^p(D-2p-1)(D-2)}{r_+^{-(D-2p-2)}}-\frac{(D-2)^{q}(D-3)^qQ^{2q}}{r_+^{4q-D+2}} \\&+m^2\bigg(\frac{c_1\epsilon(D-2)^2}{r_+^{3-D}}+\frac{c_2\epsilon^2(D-2)^2}{r_+^{4-D}}+\frac{c_3\epsilon^3(D-2)(D-3)(D-4)}{r_+^{5-D}} \\&+\frac{c_4\epsilon^4(D-2)(D-3)(D-4)(D-5)}{r_+^{6-D}}\bigg)-2r_+^{D-2}\sqrt{B+\frac{S^2}{r_+^{2D-2}}}\bigg]. \label{31}\end{split}
    \end{eqnarray}
    So, using the above value of Hawking temperature (\ref{28}) and Eq. (\ref{31}) we can obtain the entropy in the form as follows 
    \begin{equation}
    S=2\pi (D-2)\Sigma_{D-2}\sum_{p=1}^{s}\frac{\overline{\alpha}_pk^{p-1}p}{(D-2p)r_+^{-(D-2p)}}.\label{32}
    \end{equation}
    This form of entropy shows that in massive gravity, magnetically charged Lovelock black hole in the presence of dark fluid does not obey the Hawking area theorem. This behaviour is very similar to the electrically charged black holes obtained recently in Lovelock gravity \cite{27}. 
    
    The heat capacity of the black hole is defined by the relation
    \begin{equation}
    C=\frac{dM}{dT}=\frac{dM}{dr_+}\bigg(\frac{dT}{dr_+}\bigg)^{-1}. \label{33}
    \end{equation}
    So, by using Eq. (\ref{31}) and Hawking temperature we get the heat capacity in the form
    \begin{eqnarray}\begin{split}
    C=\frac{2\pi \Sigma_{D-2}r_+^{D-1}\psi_2(r_+)}{W(r_+)\bigg(\frac{\psi_1}{dr_+}+\psi_4(r_+)\bigg)+2\psi_3(r_+)}, \label{34}\end{split}
    \end{eqnarray}
    where
     \begin{eqnarray}\begin{split}
    \psi_1(r_+)&=-\frac{(D-2)^{q-1}(D-3)^qQ^{2q}}{r_+^{4q+1}}+m^2\bigg(\frac{c_1\epsilon(D-2)}{r_+^{2}}+\frac{c_2\epsilon^2(D-2)}{r_+^{3}} \\&+\frac{c_3\epsilon^3(D-3)(D-4)}{r_+^{4}}+\frac{c_4\epsilon^4(D-5)(D-3)(D-4)}{r_+^{5}}\bigg), \label{35}\end{split}
    \end{eqnarray}
    \begin{eqnarray}\begin{split}
   \frac{\psi_1}{dr_+}&=\frac{(4q+1)(D-2)^{q-1}(D-3)^qQ^{2q}}{r_+^{4q+2}}-m^2\bigg(\frac{2c_1\epsilon(D-2)}{r_+^{3}}+\frac{3c_2\epsilon^2(D-2)}{r_+^{4}} \\&+\frac{4c_3\epsilon^3(D-3)(D-4)}{r_+^{5}}+\frac{5c_4\epsilon^4(D-5)(D-3)(D-4)}{r_+^{6}}\bigg), \label{36}\end{split}
    \end{eqnarray}
    \begin{eqnarray}\begin{split}
    \psi_2(r_+)&=\sum_{p=0}^{s}\frac{\overline{\alpha}_pk^p(D-2p-1)(D-2)}{r_+^{2p+1}}-\frac{(D-2)^{q}(D-3)^qQ^{2q}}{r_+^{4q+1}}-\frac{2}{r_+}\sqrt{B+\frac{S^2}{r_+^{2D-2}}} \\&+m^2\bigg(\frac{c_1\epsilon(D-2)^2}{r_+^{2}}+\frac{c_2\epsilon^2(D-2)^2}{r_+^{3}}+\frac{c_3\epsilon^3(D-3)(D-4)(D-2)}{r_+^{4}} \\&+\frac{c_4\epsilon^4(D-2)(D-5)(D-3)(D-4)}{r_+^{5}}\bigg), \label{37}\end{split}
    \end{eqnarray}
    \begin{eqnarray}\begin{split}
    \psi_3(r_+)&=\sum_{p=0}^{s}\frac{\overline{\alpha}_pp^2k^{p-1}}{r_+^{2p+1}}\bigg[\psi_1(r_+)-\frac{2}{(D-2)r_+}\sqrt{B+\frac{S^2}{r_+^{2D-2}}}+\sum_{p=0}^{s}\frac{\overline{\alpha}_pk^p(D-2p-1)}{r_+^{2p+1}}\bigg], \label{38}\end{split}
    \end{eqnarray}
    \begin{eqnarray}\begin{split}
    \psi_4(r_+)&=\frac{2}{(D-2)r_+^2}\sqrt{B+\frac{S^2}{r_+^{2D-2}}}+\frac{2(D-1)S^2}{(D-2)r_+^{2D}}\bigg(\sqrt{B+\frac{S^2}{r_+^{2D-2}}}\bigg)^{-1} \\&-\sum_{p=0}^{s}\frac{\overline{\alpha}_pk^p(D-2p-1)(2p+1)}{r_+^{2p+2}}. \label{39}\end{split}
    \end{eqnarray} 
    
Note that the heat capacity is significant in the sense that  thermal stability of black holes is dependent on the sign of heat capacity. The region where this quantity is positive indicates that the black hole is stable whereas negativity of heat capacity corresponds to the unstable thermodynamic system. The point at which Hawking temperature and heat capacity change signs indicates the first order phase transition while the point at which heat capacity is not convergent corresponds to the second order phase transition of black hole.

\section{Magnetized Einsteinian black holes of massive gravity} 

Here we determine the metric function corresponding to magnetized black holes of massive gravity in diverse dimensions where the higher curvature terms are neglected by considering the coefficients $\alpha_p=0$ for $p\geq 2$ in the action function (\ref{1}). Note that, these black holes are also surrounded by dark fluid and the source of gravity is again the power-Yang-Mills field. Therefore, with the use of this assumption we obtain the metric function for $k=1$ from the polynomial equation (\ref{25}) as
      \begin{eqnarray}\begin{split}
      f(r)&=1-\frac{16\pi \overline{M}}{(D-2)\Sigma_{D-2}r^{D-3}}-\frac{(D-2)^{q-1}(D-3)^qQ^{2q}}{(D-4q-1)r^{4q-2}}+\frac{2r^{3-D}}{(D-2)(D-1)} \\&\times\bigg(S\arcsin{\bigg(\frac{S}{\sqrt{B}r^{D-1}}\bigg)}-r^{D-1}\sqrt{B+\frac{S^2}{r^{2D-2}}}\bigg)+\frac{\alpha_0r^2}{(D-1)(D-2)} \\&+m^2\bigg(c_1\epsilon r+\frac{c_2\epsilon^2(D-2)}{(D-3)}+\frac{c_3\epsilon^3(D-3)}{r}+\frac{c_4\epsilon^4(D-3)(D-4)}{r^2}\bigg), \label{40}\end{split}
      \end{eqnarray}
      where we have used the value $M=8\pi \overline{M}$. Choosing $D=4$ and $q=1$, the above metric function becomes
      \begin{eqnarray}\begin{split}
      f(r)&=1-\frac{2 \overline{M}}{r}-\frac{Q^{2}}{r^{2}}+\frac{1}{3r}\bigg[S\arcsin{\bigg(\frac{S}{\sqrt{B}r^{3}}\bigg)}-r^{3}\sqrt{B+\frac{S^2}{r^{6}}}\bigg] \\&+\frac{\alpha_0r^2}{6}+m^2\bigg[c_1\epsilon r+2c_2\epsilon^2+\frac{c_3\epsilon^3}{r}\bigg], \label{41}\end{split}
      \end{eqnarray}  
which is now the metric function for a massive anti-de Sitter black hole of Einstein-Yang-Mills gravity surrounded by dark fluid which satisfies a non-linear equation of state. If we choose $B=S=\alpha_0=0$ and neglect the contributions from massive gravity, then the metric function obtained corresponds to the familiar Reissner-Nordstrom like black hole with Yang-Mills magnetic charge. Eq. (\ref{40}) shows that the resulting solution has the behaviour of a non-asymptotically flat spacetime, however, by choosing $\alpha_0=2\sqrt{B}$, the solution represents an asymptotically flat spacetime.

In general, the Ricci scalar and Kretschmann scalar for the line element (\ref{6}) are given by

\begin{eqnarray}\begin{split}
R&=\bigg[(D-2)(D-3)\bigg(\frac{1-f(r)}{r^2}\bigg)-\frac{d^2f}{dr^2}-\frac{2(D-2)}{r}\frac{df}{dr}\bigg],\label{41a}\end{split}
\end{eqnarray}
and
\begin{eqnarray}\begin{split}
K&=\bigg[2(D-2)(D-3)\bigg(\frac{1-f(r)}{r^2}\bigg)^2-\bigg(\frac{d^2f}{dr^2}\bigg)^2+\frac{2(D-2)}{r^2}\bigg(\frac{df}{dr}\bigg)^2\bigg].\label{41b}\end{split}
\end{eqnarray}
So, by using the metric function (\ref{40}), one can clearly see that both Ricci and Kretschmann scalars diverge at $r=0$, which confirms that the resulting gravitating object is basically a black hole.
In $D$ dimensions the mass of the black hole in terms of the event horizon $r_+$ is given by 
    \begin{eqnarray}\begin{split}
     M&= \frac{\Sigma_{D-2}}{2}\bigg[\frac{\alpha_0r_+^{D-1}}{(D-1)}+(D-2)r_+^{D-3}-\frac{(D-2)^{q}(D-3)^qQ^{2q}}{(D-4q-1)r_+^{4q-D+1}}+m^2\bigg(\frac{c_1\epsilon(D-2)}{r_+^{2-D}}\\&+\frac{c_2\epsilon^2(D-2)^2}{(D-3)r_+^{3-D}}+\frac{c_3\epsilon^3(D-2)(D-3)}{r_+^{4-D}}+\frac{c_4\epsilon^4(D-2)(D-3)(D-4)}{r_+^{5-D}}\bigg)\\&-\frac{2}{D-1}\bigg(r_+^{D-1}\sqrt{B+\frac{S^2}{r_+^{2D-2}}}-S\arcsin{\bigg(\frac{S}{\sqrt{B}r_+^{D-1}}\bigg)}\bigg)\bigg].  \label{42}\end{split}
          \end{eqnarray}
          
\begin{figure}[h]
    	\centering
    	\includegraphics[width=0.8\textwidth]{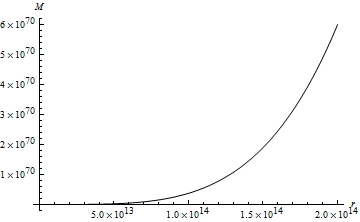}
    	\caption{Plot of function $M$ (Eq. (\ref{42})) vs $r_+$ for fixed values of $D=6$, $m=10$, $Q=100$, $B=100$, $\Sigma_{D-2}=50$, $S=500$, $c_1=10$, $c_2=20$, $c_3=30$, $c_4=40$, $\epsilon=1$, $\alpha_0=2\times 10^{14}$, $\overline{\alpha}_2=1\times10^{-14}$ and $q=2$.}\label{skr1}
    \end{figure}

\begin{figure}[h]
    	\centering
    	\includegraphics[width=0.8\textwidth]{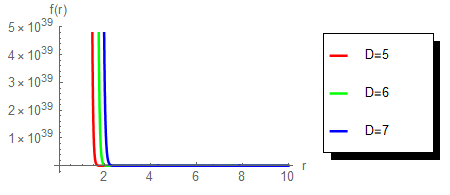}
    	\caption{Plot of function $f(r)$ (Eq. (\ref{40})) in small domain of $r$ for fixed values of $m=100$, $\overline{M}=100$, $\Sigma_{D-2}=1$, $Q=100$, $B=100$, $S=500$, $c_1=10$, $c_2=20$, $c_3=30$, $c_4=40$, $\epsilon=1$ and $q=10$.} \label{Askar1a}
    \end{figure}

\begin{figure}[h]
	\centering
	\includegraphics[width=0.8\textwidth]{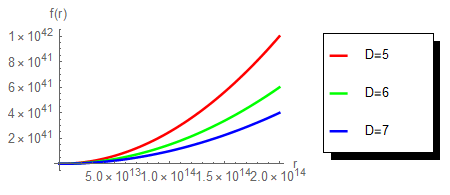}
	\caption{Graph of function $f(r)$ (Eq. (\ref{40})) in large domain of $r$ for fixed values of $m=100$, $\overline{M}=100$, $\Sigma_{D-2}=1$, $Q=100$, $B=100$, $S=500$, $c_1=10$, $c_2=20$, $c_3=30$, $c_4=40$, $\epsilon=1$ and $q=10$.} \label{Askar1b}
\end{figure}

     The plot of mass in terms of horizon radius $r_+$ is shown in Fig. (\ref{skr1}) which shows that the mass is an increasing function of $r_+$ and for all values of $M$, there exist horizons of the black hole. Fig. (\ref{Askar1a}) shows the plot of the metric function in small domain of $r$ for different values of $D$. The point where the curve touches the horizontal axis indicates position of the black hole's horizon radius. Fig. (\ref{Askar1b}) is the graph of $f(r)$ for large domain of $r$ in different dimensions, from which it becomes clear that for large values of $r$, the metric function also grows large, and thus the spacetime is non-asymptotically flat.
   Using (\ref{41}), we obtain the Hawking temperature $T=\kappa/2\pi$ in this case also in the following form  
    \begin{eqnarray}\begin{split}
    T&=\frac{r_+}{4\pi}\bigg[\frac{\alpha_0}{(D-2)}+\frac{(D-3)}{r_+^2}-\frac{(D-2)^{q-1}(D-3)^qQ^{2q}}{r_+^{4q-2}}-\frac{2}{(D-2)}\sqrt{B+\frac{S^2}{r_+^{2D-2}}}\\&+m^2\bigg(\frac{c_1\epsilon(D-2)}{r_+}+\frac{c_2\epsilon^2 (D-2)}{r_+^2}+\frac{c_3\epsilon^3 (D-3)(D-4)}{r_+^3}+\frac{c_4\epsilon^4 (D-3)(D-4)(D-5)}{r_+^4}\bigg)\bigg]. \label{43}\end{split}
    \end{eqnarray} 
    

    \begin{figure}[h]
    	\centering
    	\includegraphics[width=0.8\textwidth]{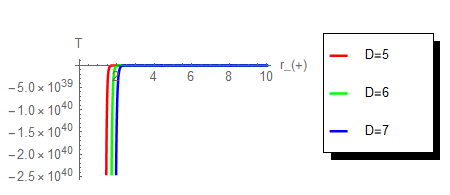}
    	\caption{Plot of temperature $T$ (Eq. (\ref{43})) vs $r_+$ in small domain of $r_+$ for fixed values of $m=100$, $Q=100$, $B=100$, $S=500$, $c_1=10$, $c_2=20$, $c_3=30$, $c_4=40$, $\epsilon=1$ and $q=10$.}\label{Saif1a}
    \end{figure} 
 \begin{figure}[h]
	\centering
	\includegraphics[width=0.8\textwidth]{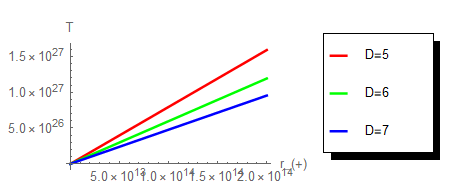}
	\caption{Temperature $T$ (Eq. (\ref{43})) vs $r_+$ in large domain of $r_+$ afor fixed values of $m=100$, $Q=100$, $B=100$, $S=500$, $c_1=10$, $c_2=20$, $c_3=30$, $c_4=40$, $\epsilon=1$ and $q=10$.}\label{Saif1b}
\end{figure}

Fig. (\ref{Saif1a}) shows the behaviour of Hawking temperature of Einsteinian black holes for different values of $D$ in small domain of $r_+$. The point at which the curve intersects $r_+$-axis indicates the first order phase transition of black hole and the region where temperature is negative shows that the black hole is unstable. Fig. (\ref{Saif1b}) is the graph of Hawking temperature in large domain of $r_+$, which shows that black holes with larger horizon radii have greater values of Hawking temperature. Now, by differentiating (\ref{42}) with respect to $r_+$ we have
     \begin{eqnarray}\begin{split}
     \frac{dM}{dr_+}&=\frac{\Sigma_{D-2}}{2}\bigg[\alpha_0r_+^{D-2}-2r_+^{D-2}\sqrt{B+\frac{S^2}{r_+^{2D-2}}}+(D-2)(D-3)r_+^{D-4}+Z_1(r_+)\bigg], \label{44}\end{split}
     \end{eqnarray} 
     where
     \begin{eqnarray}\begin{split}
     Z_1(r_+)&=m^2\bigg(c_1\epsilon(D-2)^2r_+^{D-3}+c_2\epsilon^2(D-2)^2r_+^{D-4}+c_3\epsilon^3(D-2)(D-3)(D-4)r_+^{D-5}\\&+c_4\epsilon^4(D-2)(D-3)(D-4)(D-5)r_+^{D-6}\bigg)-\frac{(D-2)^q(D-3)^qQ^{2q}}{r_+^{4q-D+2}}. \label{45}\end{split}
     \end{eqnarray}
     So, by using (\ref{44}) and the expression of Hawking temperature (\ref{43}) we get the entropy in the form
      \begin{equation}
      S=2\pi \Sigma_{D-2}r_+^{D-2},\label{46}
      \end{equation}
     which clearly indicates that unlike the case of Lovelock black holes, here entropy satisfies the area law. Differentiation of Hawking temperature yields
     \begin{eqnarray}\begin{split}
     \frac{dT}{dr_+}&=\frac{1}{4\pi}\bigg[\frac{\alpha_0}{(D-2)}+\frac{3-D}{r_+^2}-\frac{2}{D-2}\sqrt{B+\frac{S^2}{r_+^{2D-2}}}+\frac{2S^2(D-1)}{(D-2)r_+^{2D-2}\sqrt{B+\frac{S^2}{r_+^{2D-2}}}} \\&+\frac{(4q-1)(D-2)^{q-1}(D-3)^qQ^{2q}}{r_+^{4q}}-m^2\bigg(\frac{c_2\epsilon^2(D-2)}{r_+^2}+\frac{2c_3\epsilon^3(D-2)(D-4)}{r_+^3} \\&+\frac{3c_4\epsilon^4(D-3)(D-4)(D-5)}{r_+^4}\bigg)\bigg].  \label{47a}\end{split}
     \end{eqnarray}
     The heat capacity can be obtained from the general formula (\ref{33}), by using Eqs. (\ref{43})-(\ref{47a}), in the following form
     \begin{eqnarray}\begin{split}
    C&=\frac{2\pi (D-2)r_+^2}{\Psi(r_+)}\bigg[\alpha_0r_+^{D-2}-2r_+^{D-2}\sqrt{B+\frac{S^2}{r_+^{2D-2}}}+(D-3)(D-2)r_++Z_1(r_+)\bigg], \label{47}\end{split}
     \end{eqnarray}
     where
     \begin{eqnarray}\begin{split}
     \Psi(r_+)&=\alpha_0(D-2)(3-D)r_+^2-2r_+^2\sqrt{B+\frac{S^2}{r_+^{2D-2}}}\\&+\frac{2S^2(D-1)r_+^{4-2D}}{\sqrt{B+\frac{S^2}{r_+^{2D-2}}}}+(D-2)r_+^2Z_2(r_+), \label{48}\end{split}
     \end{eqnarray}
     and
     \begin{eqnarray}\begin{split}
     Z_2(r_+)&=\frac{(4q-1)(D-2)^{q-1}(D-3)^qQ^{2q}}{r_+^{4q}}-m^2\bigg[\frac{c_2\epsilon^2(D-2)}{r_+^2}\\&+\frac{2c_3\epsilon^3(D-2)(D-4)}{r_+^3}+\frac{3c_4\epsilon^4(D-3)(D-4)(D-5)}{r_+^4}\bigg]. \label{49}\end{split}
     \end{eqnarray}
\begin{figure}[h]
	\centering
	\includegraphics[width=0.8\textwidth]{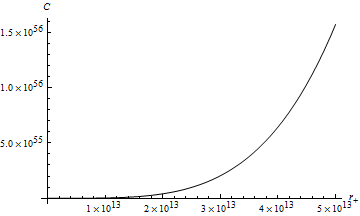}
	\caption{Plot of function $C$ (Eq. (\ref{47})) vs $r$ for fixed values of $m=100$, $D=6$, $Q=100$, $B=100$, $S=500$, $c_1=10$, $c_2=20$, $c_3=30$, $c_4=40$, $\epsilon=1$ and $q=10$.}\label{khan1}
\end{figure}
\begin{figure}[h]
	\centering
	\includegraphics[width=0.8\textwidth]{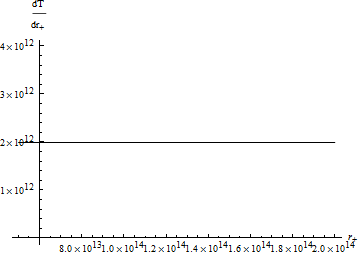}
	\caption{Graph of $dT/dr_+$ (\ref{47a}) vs $r$ for fixed values of $m=100$, $D=6$, $Q=100$, $B=100$, $S=500$, $c_1=10$, $c_2=20$, $c_3=30$, $c_4=40$, $\epsilon=1$ and $q=10$.}\label{phase1}
\end{figure}
The graph of heat capacity is plotted in Fig. (\ref{khan1}). Note that the black hole is stable in the region where both Hawking temperature and heat capacity are positive. However, Fig. (\ref{phase1}) shows that in the indicated region of $r_+$ the second order phase transition of black hole does not occur because in that domain $dT/dr_+$ can never be zero and hence heat capacity is also well defined.

\section{Magnetized Gauss-Bonnet black holes of massive gravity} 

  In this section we will work out magnetized black hole solutions of Gauss-Bonnet massive gravity. For this we consider $\alpha_2$ to be non-zero and $\alpha_p=0$ for $p\geq 3$ in the action function (\ref{1}) and choose $k=1$ in the polynomial equation (\ref{25}) to obtain the expression for the metric function in two branches as
  \begin{eqnarray}\begin{split}
  f_{\pm}(r)&=1+\frac{r^2}{2\overline{\alpha}_2^2}\bigg(1\pm\sqrt{H(r)}\bigg), \label{50}\end{split}
  \end{eqnarray}
  where $H(r)$ is given by
  \begin{eqnarray}\begin{split}
  H(r)&=1-\frac{4\alpha_0\overline{\alpha}_2}{(D-1)(D-2)}+\frac{8\overline{\alpha}_2M}{\Sigma_{D-2}(D-2)r^{D-1}}+\frac{4\overline{\alpha}_2Q^{2q}(D-2)^{q-1}(D-3)^q}{(D-4q-1)r^4q} \\&-4m^2\overline{\alpha}_2\bigg(\frac{c_1\epsilon}{r}+\frac{c_2\epsilon^2(D-2)}{(D-3)r^2}+\frac{c_3\epsilon^3(D-3)}{r^3}+\frac{c_4\epsilon^4(D-3)(D-4)}{r^4}\bigg) \\&+\frac{8\overline{\alpha}_2}{(D-2)(D-1)r^{D-1}}\bigg(r^{D-1}\sqrt{B+\frac{S^2}{r^{2D-2}}}-S\arcsin{\bigg(\frac{S}{\sqrt{B}r^{D-1}}\bigg)}\bigg). \label{51}\end{split}
  \end{eqnarray}
  The asymptotic value of $f_{\pm}(r)$ in the limit $r\rightarrow\infty$ is given by
  \begin{eqnarray}\begin{split}
  f_{\pm}(r)&=1+\frac{r^2}{2\overline{\alpha}_2}\bigg(1\pm\sqrt{1-\frac{4\overline{\alpha}_2(\alpha_0-2\sqrt{B})}{(D-1)(D-2)}}\bigg). \label{52}\end{split}
  \end{eqnarray}
  This expression shows that for any positive value of $\overline{\alpha}_2$ the function $f_+(r)$ represents anti-de Sitter spacetime for $\alpha_0>2\sqrt{B}$ and de Sitter for $\alpha_0<2\sqrt{B}$. For negative values of the coefficient $\overline{\alpha}_2$, this positive branch represents de Sitter spacetime for $\alpha_0<2\sqrt{B}$, and for the range $\alpha_0>2\sqrt{B}$ it yields anti-de Sitter spacetime. Similarly for positive values of $\overline{\alpha}_2$, the negative branch $f_-(r)$represents de Sitter for $\alpha_0>2\sqrt{B}$ and we get anti-de Sitter for $\alpha_0<2\sqrt{B}$. However, for negative values of $\overline{\alpha}_2$, the negative branch yields anti-de Sitter for $\alpha_0>2\sqrt{B}$ and de Sitter for $\alpha_0<2\sqrt{B}$. We choose the physically meaningful metric which approaches to the well known Schwarzschild metric of Einstein's theory when the effects of the cosmological constant, massive parameter $m$ and other gravitational sources are absent and there is only a central mass, then the positive branch $f_+(r)$ is not workable. Because the metric solution needs to be real-valued, we must take those values of $r$ for which the function $H(r)$ in Eq. (51) is non-negative. Now it is possible to find the root $r_b$ of equation $H(r)=0$ for given values of $M$, $Q$ and $B$. This root $r=r_b$ is called a branch singularity \cite{49,50} which makes the metric function $f_-(r)$ well defined in the interval $[r_+,\infty)$ because $H(r_{\pm})=(1+2\overline{\alpha}_2/r_{\pm}^2)^2$. This leads us to the observation that there exists branch singularity $r_b$ for the metric function $f_-(r)$ having inner horizon $r_-$ and outer horizon $r_+$, for which the inequality $r_b<r_-\leq r_+$ is satisfied.
  
  Like the previous case of Einsteinian black holes, here also one can easily verify that both the curvature invariants have singularity at $r=0$. This can be done by using the metric function (\ref{50}) in Eqs. (\ref{41a})-(\ref{41b}). Thus, we conclude that our resulting solution (\ref{50}) of Gauss-Bonnet massive gravity also describes black hole. The mass of this type of black hole in terms of horizon radius is given by 
  
  \begin{eqnarray}\begin{split}
  M&= \frac{\Sigma_{D-2}}{2}\bigg[\frac{\overline{\alpha}_2(D-2)}{r_+^{5-D}}+\frac{\alpha_0r_+^{D-1}}{(D-1)}+(D-2)r_+^{D-3}-\frac{(D-2)^{q}(D-3)^qQ^{2q}}{(D-4q-1)r_+^{4q-D+1}} \\&+m^2\bigg(\frac{c_1\epsilon(D-2)}{r_+^{2-D}}+\frac{c_2\epsilon^2(D-2)^2}{(D-3)r_+^{3-D}}+\frac{c_3\epsilon^3(D-2)(D-3)}{r_+^{4-D}} \\&+\frac{c_4\epsilon^4(D-2)(D-3)(D-4)}{r_+^{5-D}}\bigg)-\frac{2}{D-1}\bigg(r_+^{D-1}\sqrt{B+\frac{S^2}{r_+^{2D-2}}}-S\arcsin{\bigg(\frac{S}{\sqrt{B}r_+^{D-1}}\bigg)}\bigg)\bigg].\label{53}\end{split}
  \end{eqnarray} 
  
\begin{figure}[h]
  	\centering
  	\includegraphics[width=0.8\textwidth]{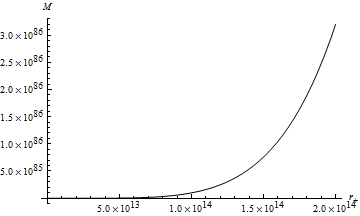}
  	\caption{Graph of $M$ (Eq. (\ref{53})) vs $r_+$ for fixed values of $D=6$, $m=10$, $Q=100$, $B=100$, $\Sigma_{D-2}=50$, $S=500$, $c_1=10$, $c_2=20$, $c_3=30$, $c_4=40$, $\epsilon=1$, $\alpha_0=2\times 10^{14}$, $\overline{\alpha}_2=1\times10^{-14}$ and $q=2$.}\label{skr2}
  \end{figure}

\begin{figure}[h]
  	\centering
  	\includegraphics[width=0.8\textwidth]{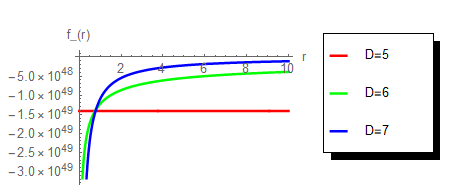}
  	\caption{Plot of function $f(r)$ (Eq. (\ref{50})) in small domain for fixed values of $M=1.5\times10^{86}$, $m=10$, $Q=100$, $B=100$, $\Sigma_{D-2}=50$, $S=500$, $c_1=10$, $c_2=20$, $c_3=30$, $c_4=40$, $\epsilon=1$, $\alpha_0=2\times 10^{14}$, $\overline{\alpha}_2=1\times10^{-14}$ and $q=2$.}\label{Askar2a}
  \end{figure}

\begin{figure}[h]
	\centering
	\includegraphics[width=0.8\textwidth]{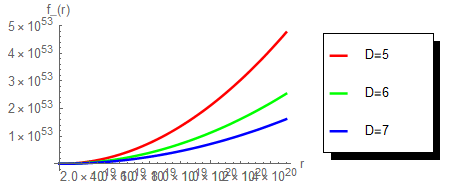}
	\caption{Plot of function $f(r)$ (Eq. (\ref{50})) in large domain for fixed values of $M=1.5\times10^{86}$, $m=10$, $Q=100$, $B=100$, $\Sigma_{D-2}=50$, $S=500$, $c_1=10$, $c_2=20$, $c_3=30$, $c_4=40$, $\epsilon=1$, $\alpha_0=2\times 10^{14}$, $\overline{\alpha}_2=1\times10^{-14}$ and $q=2$.}\label{Askar2b}
\end{figure}

  Fig. (\ref{skr2}) is graph of the mass function in terms of the horizon radius showing that $M$ is an increasing function of $r_+$ and for any value in the range of $M$ there exists horizon of the black hole. The metric function $f(r)$ is plotted in Fig. (\ref{Askar2a}) for small domain and in Fig. (\ref{Askar2b}) for large domain; the point at which the indicated curve intersects the horizontal axis gives location of the event horizon for the given fixed values of parameters in the metric function. Similarly, the Hawking temperature of the obtained Gauss-Bonnet black hole is given by
  \begin{eqnarray}\begin{split}
  T&=\frac{r_+^2}{4\pi \bigg(1+\frac{2\overline{\alpha}_2}{r_+^2}\bigg)}\bigg[\frac{\overline{\alpha}_2(D-1)}{r_+^5}+\frac{(D-1)}{r_+^2}+\frac{\alpha_0}{(D-2)r_+}-\frac{2}{(D-2)r_+} \\&\times\sqrt{B+\frac{S^2}{r^{2D-2}_+}}+m^2\bigg(\frac{c_1\epsilon (D-2)}{r_+^2}+\frac{c_2\epsilon^2 (D-2)}{r_+^3}+\frac{c_3\epsilon^3 (D-3)(D-4)}{r_+^4} \\&+\frac{c_4\epsilon^4 (D-3)(D-4)(D-5)}{r_+^5}\bigg)-\frac{(D-2)^{q-1}(D-3)^qQ^{2q}}{r_+^{1+4q}}\bigg]-\frac{1}{2\pi r_+}.\label{54}\end{split}
  \end{eqnarray} 
  

\begin{figure}[h]
  	\centering
  	\includegraphics[width=0.8\textwidth]{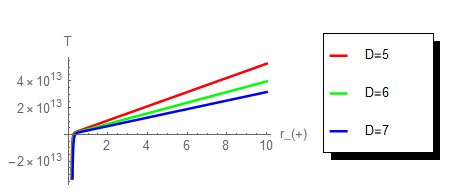}
  	\caption{Graph of temperature $T$ (Eq. (\ref{54})) vs $r_+$ in small domain for fixed values of $m=10$, $Q=100$, $B=100$, $\Sigma_{D-2}=50$, $S=500$, $c_1=10$, $c_2=20$, $c_3=30$, $c_4=40$, $\epsilon=1$, $\alpha_0=2\times 10^{14}$, $\overline{\alpha}_2=1\times10^{-14}$ and $q=2$.}\label{Saif2a}
  \end{figure}

\begin{figure}[h]
	\centering
	\includegraphics[width=0.8\textwidth]{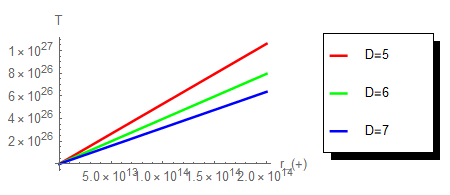}
	\caption{Temperature $T$ (Eq. (\ref{54})) vs $r_+$ in large domain for fixed values of $m=10$, $Q=100$, $B=100$, $\Sigma_{D-2}=50$, $S=500$, $c_1=10$, $c_2=20$, $c_3=30$, $c_4=40$, $\epsilon=1$, $\alpha_0=2\times 10^{14}$, $\overline{\alpha}_2=1\times10^{-14}$ and $q=2$.}\label{Saif2b}
\end{figure}

  The behaviour of Hawking temperature of Gauss-Bonnet black holes for different values of $D$ is shown in Fig. (\ref{Saif2a}) where the domain of definition has been taken very small. This graph shows that first order phase transitions occur for these types of black holes since the temperature clearly changes sign in the indicated domain, for instance, at $r_+=0.4$ in six dimesnions. The black hole is unstable in that region where the range of temperature curve contains negative values. The behaviour of temperature in large domain is shown in Fig. (\ref{Saif2b}). Since the range of $T$ is the set of positive real numbers for all values of $r_+>r_c$, so we conclude that the black hole temperature is positive and increases linearly with the horizon radius for all these values. Here $r_c$ represents the point at which Hawking temperature changes sign, for example, $r_c=0.4$ in six dimensional spacetime.   
  
  Now, differentiating (\ref{53}) with respect to $r_+$ we have
 \begin{eqnarray}\begin{split}
 \frac{dM}{dr_+}&= \frac{\Sigma_{D-2}}{2}\bigg[\overline{\alpha}_2(D-2)(D-5)r_+^{D-6}+(D-2)(D-3)r_+^{D-4}-\frac{2}{r_+^{2-D}}\sqrt{B+\frac{S^2}{r_+^{2D-2}}} \\&+m^2\bigg(\frac{c_1\epsilon(D-2)^2}{r_+^{3-D}}+\frac{c_2\epsilon^2(D-2)^2}{r_+^{4-D}}+\frac{c_3\epsilon^3(D-2)(D-3)(D-4)}{r_+^{5-D}} \\&+\frac{c_4\epsilon^4(D-2)(D-3)(D-4)(D-5)}{r_+^{6-D}}\bigg)-\frac{Q^{2q}(D-2)^q(D-3)^q}{r_+^{4q+2-D}}\bigg],\label{55}\end{split}
 \end{eqnarray}
 So, by using Eqs. (\ref{54}) and (\ref{55}) in (\ref{30}) we get entropy for the Gauss-Bonnet black hole as
 \begin{equation}
 S=2(D-2)\pi\Sigma_{D-2}r_+^{D-4}\bigg(\frac{r^2_+}{D-2}+\frac{2 \overline{\alpha}_2}{D-4}\bigg),
 \label{56}
 \end{equation}
 which clearly shows that in this case also the area law is not satisfied (as is the case in Lovelock black holes). Differentiation of (\ref{54}) gives
  \begin{eqnarray}\begin{split}
  \frac{dT}{dr_+}&= \frac{r_+^2}{4\pi\bigg(1+\frac{2\overline{\alpha}_2}{r_+^2}\bigg)}\frac{d\chi_1}{dr_+}+\frac{r_+\chi_1(r_+)}{2\pi\bigg(1+\frac{2\overline{\alpha}_2}{r_+^2}\bigg)}+\frac{1}{4\pi r_+^2}+\frac{\overline{\alpha}_2\chi_1(r_+)}{\pi r_+\bigg(1+\frac{2\overline{\alpha}_2}{r_+^2}\bigg)^2},\label{57}\end{split}
  \end{eqnarray}
  where
  \begin{eqnarray}\begin{split}
  \chi_1(r_+)&=\frac{\overline{\alpha}_2(D-1)}{r_+^5}+\frac{D-1}{r_+^2}+\frac{\alpha_0}{(D-2)r_+}-\frac{2}{(D-1)r_+}\sqrt{B+\frac{S^2}{r_+^{2D-2}}}\\&+m^2\bigg(\frac{c_1\epsilon(D-2)}{r_+^2}+\frac{c_2\epsilon^2(D-2)}{r_+^3}+\frac{c_3\epsilon^3(D-3)(D-4)}{r_+^4}\\&+\frac{c_4\epsilon^4(D-3)(D-4)(D-5)}{r_+^5}\bigg)-\frac{(D-2)^{q-1}(D-3)^qQ^{2q}}{r_+^{4q+1}},\label{58}\end{split}
  \end{eqnarray}
  \begin{eqnarray}\begin{split}
  \frac{d\chi_1}{dr_+}&=\frac{5\overline{\alpha}_2(1-D)}{r_+^6}+\frac{2(1-D)}{r_+^3}-\frac{\alpha_0}{r_+^2(D-2)}+\frac{Q^{2q}(4q+1)(D-2)^{q-1}(D-3)^q}{r_+^{4q}}\\&+m^2\bigg(\frac{2c_1\epsilon(D-2)}{r_+^3}+\frac{3c_2\epsilon^2(D-2)}{r_+^4}+\frac{5c_4\epsilon^4(D-3)(D-4)(D-5)}{r_+^6}\\&+\frac{4c_3\epsilon^3(D-3)(D-4)}{r_+^5}\bigg)+\frac{2}{(D-1)r_+^2}\sqrt{B+\frac{S^2}{r_+^{2D-2}}}+\frac{2S^2}{r_+^{2D}\sqrt{B+\frac{S^2}{r_+^{2D-2}}}}.\label{59}\end{split}
  \end{eqnarray}
 Finally, using the quantities calculated above, the heat capacity of our Gauss-Bonnet black hole can be expressed in the following form
 \begin{eqnarray}\begin{split}
 C&=\frac{2\pi \Sigma_{D-2}r_+^2\bigg(1+\frac{2\overline{\alpha}_2}{r_+^2}\bigg)^2\bigg[\chi_2(r_+)+(D-2)(D-3)r_+^{D-4}+\overline{\alpha}_2(D-2)(D-5)r_+^{D-6}\bigg]}{\bigg[r_+^4\bigg(1+\frac{2\overline{\alpha}_2}{r_+^2}\bigg)\frac{d\chi_1}{dr_+}+2r_+^3\chi_1(r_+)\bigg(1+\frac{2\overline{\alpha}_2}{r_+^2}\bigg)+\bigg(1+\frac{2\overline{\alpha}_2}{r_+^2}\bigg)^2+4\overline{\alpha}_2r_+\chi_1(r_+)\bigg]} ,\label{60}\end{split}
 \end{eqnarray}
 where 
 \begin{eqnarray}\begin{split}
 \chi_2(r_+)&=m^2\bigg(c_1\epsilon(D-2)^2r_+^{D-3}+c_2\epsilon^2(D-2)^2r_+^{D-4}+c_3\epsilon^3(D-2)(D-3)(D-4)r_+^{D-5}\\&+c_4\epsilon^4(D-2)(D-3)(D-4)(D-5)r_+^{D-6}\bigg)-\frac{2}{r_+^{2-D}}\sqrt{B+\frac{S^2}{r_+^{2D-2}}}\\&-\frac{(D-2)^q(D-3)^qQ^{2q}}{r_+^{4q+2-D}}.\label{61}\end{split}
 \end{eqnarray}
\begin{figure}[h]
	\centering
	\includegraphics[width=0.8\textwidth]{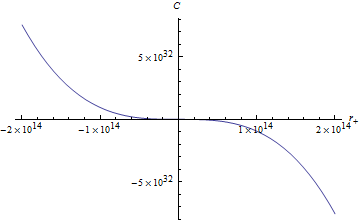}
	\caption{Plot of function $C$ (Eq. (\ref{60})) vs $r_+$ for fixed values of $D=6$, $m=10$, $Q=100$, $B=100$, $\Sigma_{D-2}=50$, $S=500$, $c_1=10$, $c_2=20$, $c_3=30$, $c_4=40$, $\epsilon=1$, $\alpha_0=2\times 10^{14}$, $\overline{\alpha}_2=1\times10^{-14}$ and $q=2$.}\label{khan2}
\end{figure}
\begin{figure}[h]
	\centering
	\includegraphics[width=0.8\textwidth]{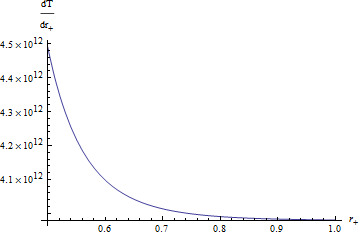}
	\caption{Plot of function $dT/dr_+$ (Eq. (\ref{57})) vs $r_+$ for fixed values of $D=6$, $m=10$, $Q=100$, $B=100$, $\Sigma_{D-2}=50$, $S=500$, $c_1=10$, $c_2=20$, $c_3=30$, $c_4=40$, $\epsilon=1$, $\alpha_0=2\times 10^{14}$, $\overline{\alpha}_2=1\times10^{-14}$ and $q=2$.}\label{phase2}
\end{figure}
The graph of heat capacity is given in Fig. (\ref{khan2}). The point at which heat capacity changes sign represents the first order phase transition of the black hole and the region in which this quantity is negative shows the instability of the hole in that region. Fig. (\ref{phase2}) shows the behaviour of $dT/dr_+$ versus $r_+$. The values of $r_+$ for which $dT/dr_+$ vanishes indicate the second order phase transition of the black hole because heat capacity is singular there. 

\section{Summary and conclusion} 

In this paper, the general static spherically-symmetric line element is assumed and massive Lovelock black holes are obtained in the presence of power-Yang-Mills field and Chaplygin-like dark fluid. After the coupling of this massive gravity with matter contents we solved the resulting gravitational field equations. We obtained a polynomial equation (\ref{25}) which generates the metric function for magnetized Lovelock black holes surrounded by dark fluid in massive gravity. We considered only pure Yang-Mills magnetic field because the problem with the electric type fields is that $\Upsilon^q$ may not be real for any q, which is not the case with pure magnetic type Yang-Mills field. We note that this is true for the power-Maxwell case as well. We then studied thermodynamics of these objects and calculated different thermodynamic quantities associated with the polynomial equation in terms of the event horizon. After this we discussed two special cases of (\ref{25}) and derived metric functions in terms of parameters $q$, $B$ and $S$ for magnetized Einsteinian and Gauss-Bonnet black holes in massive gravity. It is shown that the resulting black holes are non-asymptotically flat and the associated spacetime geometries have true curvature singularities at $r=0$. In both the cases, thermodynamic properties are studied and quantities such as mass, Hawking temperature and heat capacity are calculated and plotted for fixed values of different parameters arising in the corresponding metric functions. We discussed the behaviour of these quantities and pointed out regions where the heat capacity and Hawking temperature are positive which indicate the black hole's thermal stability. Furthermore, it is also shown that phase transitions are also possible for the black hole solutions that we have obtained. The existence of the first order phase transition is associated with the value of $r_+$ at which either the temperature or heat capacity or both change signs, while the second order phase transitions of black holes are related to the zeros of $dT/dr_+=0$, or to the points at which the heat capacity diverges. 
  
 It should be noted that when $q=1$ in each case, the corresponding metric functions for black holes surrounded by dark fluid with the standard Yang-Mills source can be obtained, but for $D=4q+1$ the solutions do not exist. For $B=S=0$ we get the solutions corresponding to Lovelock, Einsteinian and Gauss-Bonnet black holes of massive gravity, where only the power-Yang-Mills field is the gravitational source. By choosing  $Q=0$ the corresponding neutral black holes in such gravities are obtained. It should also be noted that the obtained Einsteinian black hole solutions are valid even for $q=D/4$ unlike the black holes of Einstein gravity in the presence of only Yang-Mills or power-Yang-Mills sources. This behaviour is due to the presence of massive gravity coupled to the power-Yang-Mills field which does not make the scalar curvature zero even when the trace of the energy-momentum tensor is zero at $q=D/4$. It is also important to note that when $m=B=S=0$ then our Einsteinian black hole solutions are valid only for $D\geq5$. However, our Lovelock and Gauss-Bonnet black hole solutions are valid in any dimensions such that $D\neq4q+1$ even when $m=B=S=0$. Furthermore, in case the contributions of massive gravity are neglected then the solutions of this paper, in all the cases, produce new classes of charged black holes in the presence of dark fluid. 
 
It will also be very interesting to study these types of Lovelock black holes of massive gravity in the presence of Lorentz symmetry breaking and corresponding hairy black holes of this theory. Also, it will be worthwhile to study the causal structure and causality conditions of the black hole solutions obtained in this paper. In addition to this, the causal structure of massive Lovelock gravity in the cosmological framework should also be studied.

\section*{Acknowledgements}
Research grant from the Higher Education Commission of Pakistan under its Project No. 6151 is gratefully acknowledged.

\end{document}